\documentclass[letter,twocolumn]{jpsj2} 
\def\runauthor{Shin-ichi \textsc{Kimura} \textit{et al}.}
\title{
Origin of Middle-Infrared Peaks in Cerium Compounds
}
\author{
Shin-ichi \textsc{Kimura}$^{1,2,}$\thanks{E-mail address: kimura@ims.ac.jp}, 
Takuya \textsc{Iizuka}$^{2}$,
and
Yong-seung \textsc{Kwon}$^{3}$
}

\inst{
$^{1}$UVSOR Facility, Institute for Molecular Science, Okazaki 444-8585, Japan\\
$^{2}$School of Physical Sciences, The Graduate University for Advanced Studies (SOKENDAI), Okazaki 444-8585, Japan\\
$^{3}$Department of Physics, Sungkyunkwan University, Suwon 440-746, South Korea
}

\recdate
{
\today
}

\abst{
We have demonstrated that the middle-infrared (mid-IR) peaks in the optical conductivity spectra of Ce$X_3$ ($X$~=~Pd, Sn, In) can be explained by first-principle band structure calculation with the spin-orbit interaction.
The mid-IR peak shapes in these materials are not identical to one another: CePd$_3$, CeSn$_3$, and CeIn$_3$ have a triple-peak structure, double-peak structure and broad single-peak structure, respectively.
These peaks can be theoretically explained by the optical transition from the occupied state to the spin-orbit splitted Ce~$4f$ state.
This result indicates that the mid-IR peaks originate from the simple band picture with the Ce~$4f$ state near the Fermi level, not from the conventional $cf$ hybridization gap based on the periodic Anderson model.
}
\kword{CePd$_3$, CeSn$_3$, CeIn$_3$, optical conductivity, electronic structure, $cf$ hybridization}
\begin{document}
\maketitle
%
%
Heavy fermion as well as mixed valent materials such as cerium (Ce), ytterbium (Yb), and uranium compounds commonly have a characteristic peak structure (namely, the mid-IR peak) in the middle-infrared (mid-IR) region below 1~eV in their optical conductivity [$\sigma(\omega)$] spectra.~\cite{Degiorgi1999}
The origin of the mid-IR peak has remained a long debated issue so far.
Recently, two proposals have been made in different studies regarding the origin of the mid-IR peak;
one attributing it to the optical transition between the bonding and antibonding state of the hybridization gap between conduction and local $4f$ electrons, namely $cf$ hybridization gap, predicted by the periodic Anderson model (PAM),~\cite{Dordevic2001,Hancock2004,Hancock2006,Okamura2007}
and the other to a peculiar band structure in which the $4f$ state is located near the Fermi level ($E_{\rm F}$).~\cite{Kuroiwa2007}
The former study presented a persuasive scaling plot of the energy of the mid-IR peak versus the hybridization intensity.~\cite{Okamura2007}
Actually, the $cf$ hybridization plays an important role in the physics of heavy fermion systems.~\cite{Hewson1993}
The $cf$ hybridization band predicted by PAM has been directly observed with Ce $4d-4f$ resonant angle-resolved photoemission spectroscopy of CeCoGe$_{1.2}$Si$_{0.8}$ that is a typical heavy fermion compound.~\cite{Im2008}
On the other hand, the latter study pointed out that the mid-IR peak of YbAl$_3$ can be explained in terms of a simple band picture without reference to the $cf$ hybridization gap but they shifted the Yb~$4f$ level near the Fermi level.~\cite{Kuroiwa2007} 

The mid-IR peak is mainly observed in Ce and Yb compounds.
However the shape of the mid-IR peak is different from one another.
Yb compounds have a single-peak structure due to the optical transition from the Yb $4f_{7/2}$ state because the transition from the Yb $4f_{5/2}$ state does not appear in the mid-IR region but in the near-infrared region above 1~eV due to a spin-orbit splitting (SOS) energy of about 1.4~eV.~\cite{Suga2005}
On the other hand, almost all Ce compounds have a double-peak structure with an energy split of about 0.25~eV originating from the splitting of the Ce $4f_{5/2}$ and $4f_{7/2}$ states due to the spin-orbit interaction (SOI).~\cite{Kwon2006}
For instance, CeSn$_3$ that is a typical mixed valent material has a double-peak structure.~\cite{Iizuka2008}
In CePd$_3$ that is another mixed valent material with the identical Cu$_3$Au-type crystal structure, there is a triple-peak structure with equally spaced energy splitting of 0.25~eV in the mid-IR region.~\cite{Webb1986, Kimura2003}
Another isostructural material, CeIn$_3$, which is in the antiferromagnetic ground state with the N\'eel temperature ($T_{\rm N}$) of about 10~K, also has a broad peak in the mid-IR region, not only in the paramagnetic phase above but also in the antiferromagnetic phase below $T_{\rm N}$.~\cite{Iizuka2008}
If the mid-IR peak of CeIn$_3$ originated from the $cf$ hybridization gap, CeIn$_3$ should not have a mid-IR peak below $T_{\rm N}$ because the Ruderman-Kittel-Kasuya-Yoshida (RKKY) interaction predominates over the $cf$ hybridization in the ground state. 
To clarify the origin of the different mid-IR peak shape in Ce$X_3$ ($X$~=~Pd, Sn, In) is important for the identification of the origin of the mid-IR peaks in Ce compounds.

In this Letter, to investigate the different shapes of the mid-IR peaks of Ce$X_3$ as well as the origin of the mid-IR peaks, we calculate the $\sigma(\omega)$ spectra of these materials from the first-principle band calculation with SOI.
The result shows that the mid-IR peak can be explained in terms of the optical transition derived from the band calculation.
Therefore, it is concluded that the mid-IR peaks originate from the characteristic band structure, not from the conventional $cf$ hybridization gap.

%
%
The band structure calculation was performed using the full potential linearized augmented plane wave plus local orbital (LAPW+lo) method including SOI implemented in the {\sc Wien2k} code.~\cite{WIEN2k}
Ce$X_3$ ($X$~=~Pd, Sn, In) forms a Cu$_3$Au-type cubic crystal structure ($Pm3m$, No.~221) with lattice constants of 4.1280~\AA~\cite{Buschow1979}, 4.7423~\AA~\cite{Buschow1979}, and 4.6876~\AA~\cite{Buschow1979, Buschow1969}, respectively.
The non-overlapping muffin-tin (MT) sphere radii values of 2.50 Bohr radius were used for both the Ce and $X$ atoms in Ce$X_3$.
The value of $R_{MT}K_{max}$ (the smallest MT radius multiplied by the maximum $k$ value in the expansion of plane waves in the basis set), which determines the accuracy of the basis set used, was set to 7.0.
The total number of Brillouin zones was sampled with 40,000~$k$-points.
The band structure of LaPd$_3$ (lattice constant = 4.235~\AA~\cite{Koenig1988}) was also calculated for reference.

The $\sigma(\omega)$ spectra were derived from a function as follow in which direct interband transitions were assumed;~\cite{Ant04}
\[
\hat{\sigma}(\omega) = \frac{\pi e^2}{m_0^2 \omega} \sum_{\vec{k}} \sum_{n n'} \frac{|\langle n' \vec{k}|\vec{e} \cdot \vec{p}|n \vec{k}\rangle |^{2}}{\omega - \omega_{n n'}(\vec{k})+i\Gamma} \times \frac{f(\epsilon_{n\vec{k}})-f(\epsilon_{n'\vec{k}})}{\omega_{n n'}(\vec{k})}
\]
Here, the $|n' \vec{k}\rangle$ and $|n \vec{k}\rangle$ states denote the unoccupied and occupied states, respectively, $\vec{e}$ and $\vec{p}$ are the polarization of light and the momentum of the electron, respectively, $f(\epsilon_{n\vec{k}})$ is the Fermi-Dirac distribution function, $\hbar\omega_{n n'}=\epsilon_{n\vec{k}}-\epsilon_{n'\vec{k}}$ is the energy difference of the unoccupied and occupied states and $\Gamma$ is the lifetime parameter.
In the calculation, $\Gamma$~=~1~meV was assumed.


We measured the reflectivity [$R(\omega)$] spectra of CeSn$_3$ and CeIn$_3$.
The near-normal incident optical $R(\omega)$ spectra of polycrystalline CeSn$_3$ and CeIn$_3$ were acquired in the photon energy rage of 2~meV~--~30~eV using synchrotron-based equipment at UVSOR-II.
The $\sigma(\omega)$ spectra were derived from Kramers-Kronig analysis of the $R(\omega)$ spectrum.
For CePd$_3$, we used the $\sigma(\omega)$ spectrum reported elsewhere.~\cite{Kimura2003}

%
%
\begin{figure}[t]
\begin{center}
\includegraphics[width=0.45\textwidth]{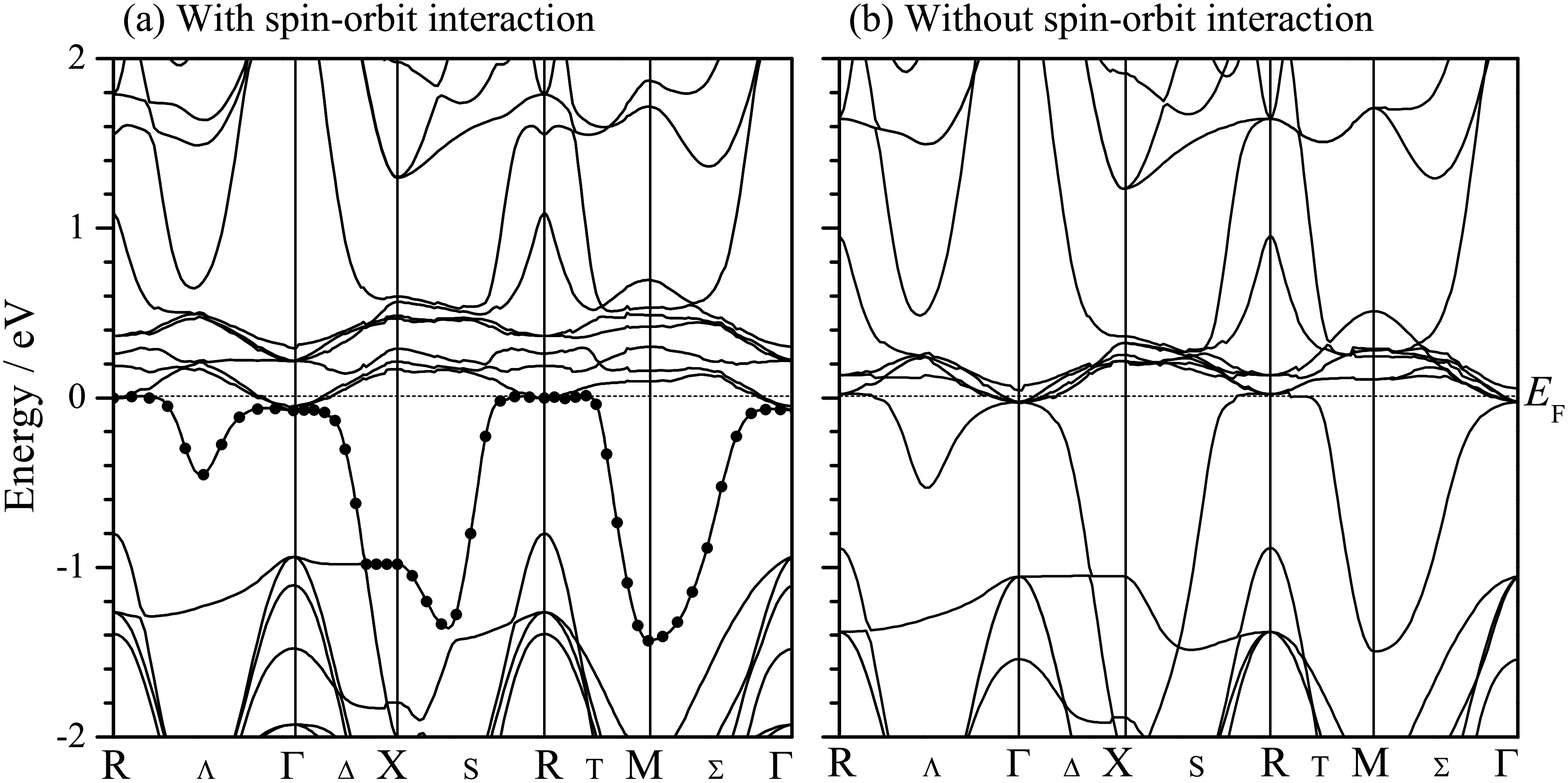}
\end{center}
\caption{
Calculated electronic structure of CePd$_3$ with (a) and without (b) the spin-orbit interaction.
The band with solid circles in (a) is the initial state of the mid-IR peaks.
}
\label{CePd3Band}
\end{figure}
%
The calculated band structures of CePd$_3$ with and without SOI are plotted in Fig.~\ref{CePd3Band}.
The highly dispersive bands below -1~eV, highly dispersive bands above -1~eV, and flat bands at 0~--~0.5~eV above $E_{\rm F}$ mainly originate from the Pd~$4d$, nearly free electron, and Ce~$4f$ states, respectively.
In Fig.~\ref{CePd3Band}(a), the Ce~$4f$ state splits into two bands with energy splitting of about 0.3~eV by SOI.
The band structure is fundamentally identical to that previously reported.~\cite{Hasegawa1987, Shekar2005}

\begin{figure}[t]
\begin{center}
\includegraphics[width=0.35\textwidth]{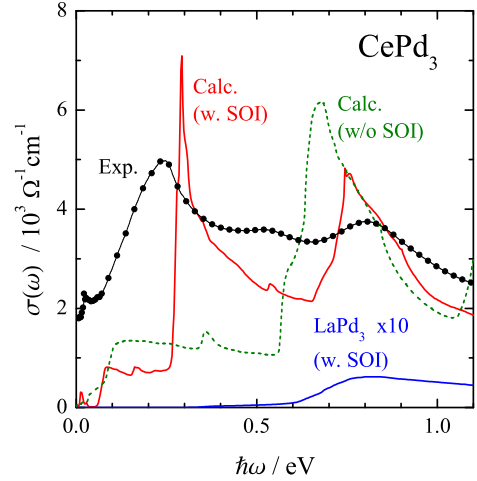}
\end{center}
\caption{
(Color online) Experimental and calculated optical conductivity [$\sigma(\omega)$] spectra of CePd$_3$.
The calculated $\sigma(\omega)$ spectrum of LaPd$_3$ is also plotted for reference.
Two different calculations with and without SOI were performed.
The experimental spectrum, which was obtained at $T$~=~8~K, is derived from that reported elsewhere.~\cite{Kimura2003}
}
\label{CePd3OC}
\end{figure}
%
The $\sigma(\omega)$ spectra calculated from the two band structures with and without SOI in Fig.~\ref{CePd3Band} are plotted in Fig.~\ref{CePd3OC}.
The experimental $\sigma(\omega)$ spectrum of CePd$_3$ at $T$~=~8~K is also plotted in the figure.
The experimental spectrum, has three large peaks at 0.25, 0.55, and 0.78~eV, and one small peak at 0.04~eV.
The calculated $\sigma(\omega)$ spectrum with SOI also has three peaks at 0.3, 0.56, and 0.75~eV and one shoulder at 0.1~eV, despite the very weak $\sigma(\omega)$ intensity of LaPd$_3$.
This structure can be assigned to the experimental structure as follows:
The peaks mainly appear in the $\Gamma$--X--R--M plane and the initial state of all peaks is commonly the highly dispersive conduction band (the line with solid circles in Fig.~\ref{CePd3Band}).
The peaks at 0.3~eV and 0.78~eV in the calculated spectrum originate from the optical transitions to the Ce~$4f_{7/2}$ state near the $\Gamma$-point and near the bottom of the $\Lambda$-axis, respectively.
The small peak at 0.55~eV originates from the transition near the $\Delta$- and S-axes.
The shoulder structure at 0.1~eV in the calculation originates from the transition from the occupied flat band to the unoccupied Ce~$4f_{5/2}$ state near the $\Gamma$-point.
Therefore, all of the mid-IR peaks of CePd$_3$ can be explained by the band structure calculation with SOI.

In the calculated spectrum without SOI, one large peak at 0.7~eV appears, as shown in Fig.~\ref{CePd3OC}.
The $\sigma(\omega)$ spectrum is fundamentally equal to the calculation reported previously.~\cite{Koenig1988}
The spectrum cannot reproduce the experimental triple-peak structure.
The SOS of the Ce~$4f$ state therefore plays an important role in the mid-IR triple-peak structure of CePd$_3$; {\it i.e.}, the mid-IR peaks originate from the band structure including the SOS of the Ce~$4f$ state.

%
\begin{figure}[t]
\begin{center}
\includegraphics[width=0.35\textwidth]{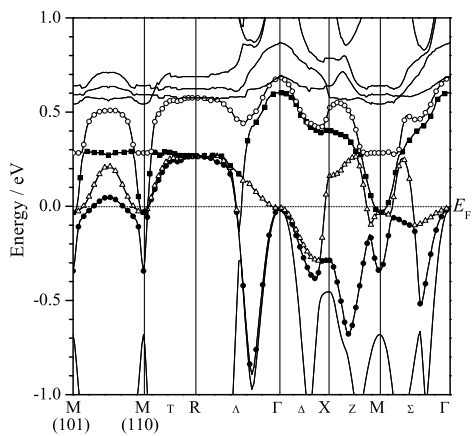}
\end{center}
\caption{
Calculated electronic structure of CeSn$_3$ with the spin-orbit interaction.
The bands with solid circles, open triangles, solid squares and open circles are the origin of the mid-IR peaks in this material.
}
\label{CeSn3Band}
\end{figure}
%
\begin{figure}[t]
\begin{center}
\includegraphics[width=0.35\textwidth]{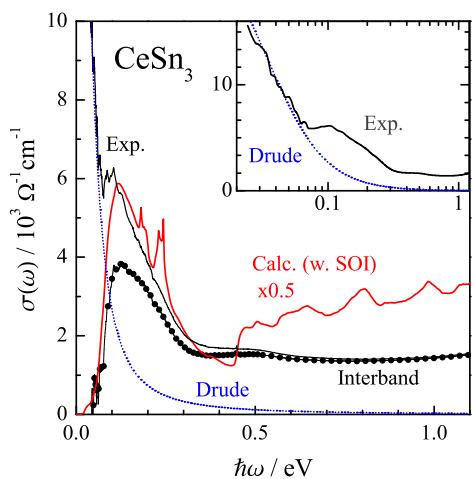}
\end{center}
\caption{
(Color online) Experimental and calculated optical conductivity [$\sigma(\omega)$] spectra of CeSn$_3$.
The calculation was performed with SOI.
The experimental spectrum was recorded at $T$~=~8~K.
The interband spectrum was obtained by subtracting the Drude component due to the carriers from the experimental $\sigma(\omega)$ spectrum.
The inset shows the lower energy part of the experimental $\sigma(\omega)$ spectrum and the Drude curve fitted at the slope at around 0.04~eV.
}
\label{CeSn3OC}
\end{figure}
To investigate the origin of the mid-IR peak in other Ce compounds, the band structure of CeSn$_3$, a mixed valent material,  was calculated as shown in Fig.~\ref{CeSn3Band} and the calculated $\sigma(\omega)$ spectrum was compared with the experimental spectrum, as shown in Fig.~\ref{CeSn3OC}.
The band structure is consistent with that previously reported by Hasegawa {\it et al.}~\cite{Hasegawa1990}
The experimental spectrum has two peaks at about 0.15 and 0.45~eV, and a large Drude structure due to the carriers.
Since the calculated $\sigma(\omega)$ spectrum only has the interband component, the experimental interband spectrum after subtracting the Drude part from the experimental $\sigma(\omega)$ spectrum is compared with the calculation.
The Drude part was evaluated by the slope of the experimental spectrum at around 0.04~eV (see the inset of Fig.~\ref{CeSn3OC}).
The calculated $\sigma(\omega)$ spectrum has a large peak at 0.15~eV, a dip at 0.4~eV, and a shoulder at 0.5~eV, and is in good agreement with the experimental spectrum.
According to the calculation for each band contribution (not shown), the peak at 0.15~eV can be assigned to originate from the transition between the similarly dispersive bands of solid circle and open triangle in Fig.~\ref{CeSn3Band} in between the M~$(101)$ and M~$(110)$-points.
The two bands expand with the parallel dispersion in a wide area near the M-point, so a large joint density of states or a $\sigma(\omega)$ peak is produced.
On the other hand, the shoulder structure at 0.5~eV originates from the transition between the bands indicated by solid circles and open circles in Fig.~\ref{CeSn3Band}.
In addition, small peaks at around 0.2~eV, which originates from the transition between the bands of solid circles and solid squares, also correspond to the shoulder at 0.2~eV in the experimental spectrum.
The experimental mid-IR peak structure of CeSn$_3$ can therefore also be explained by the band structure calculation.
However, it is noted that the intensity of the calculated $\sigma(\omega)$ spectrum is more than double the experimental value.
The same tendency has been observed in other materials, although the reason for this is not clear at present.~\cite{Kimura2007}
One possibility is the strong electron correlation in Ce~$4f$ electrons.
Since the value of the on-site Coulomb repulsion energy ($U$) is about 6~eV in Ce~$4f$ electrons, if the $cf$ hybridization intensity is not so large, some of the Ce~$4f$ states move away from $E_{\rm F}$.~\cite{Haule2005}
Therefore, the transition intensity to the Ce~$4f$ state near $E_{\rm F}$ may become smaller that the calculated value.
To confirm the electron correlation effect of the mid-IR peak, band calculation incorporating dynamical mean-field theory is needed.

%
Finally, we investigate whether the mid-IR peak of CeIn$_3$ in the antiferromagnetic ground state can be explained by the band calculation.
Generally, the band structure calculation of a material with local $4f$ electrons is performed with $U$, namely LDA+U method.~\cite{Ant04}
In the case of CeIn$_3$, however, the $U$ is considered to be small because the material is located near QCP.
Actually, the angle-resolved photoemission spectra of CeRu$_2$(Si$_{0.82}$Ge$_{0.18}$)$_2$ that is located in the slightly local regime near QCP can be explained by the itinerant band calculation.~\cite{Yamagami2008}
Since CeIn$_3$ is considered to be in the same situation as CeRu$_2$(Si$_{0.82}$Ge$_{0.18}$)$_2$, the itinerant band calculation of CeIn$_3$ was performed and compared with the experimental $\sigma(\omega)$ spectrum.

\begin{figure}[t]
\begin{center}
\includegraphics[width=0.35\textwidth]{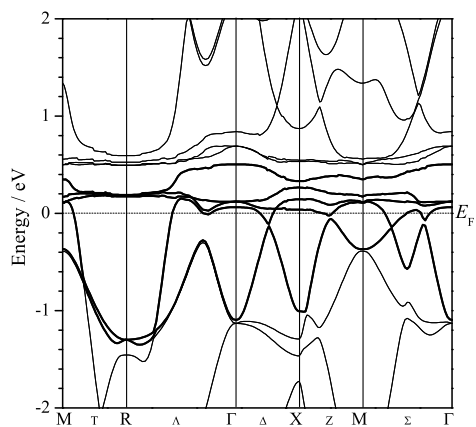}
\end{center}
\caption{
Calculated electronic structure of CeIn$_3$ with the spin-orbit interaction.
The bands with bold lines relates to the mid-IR peak of CeIn$_3$.
}
\label{CeIn3Band}
\end{figure}
%
\begin{figure}[t]
\begin{center}
\includegraphics[width=0.35\textwidth]{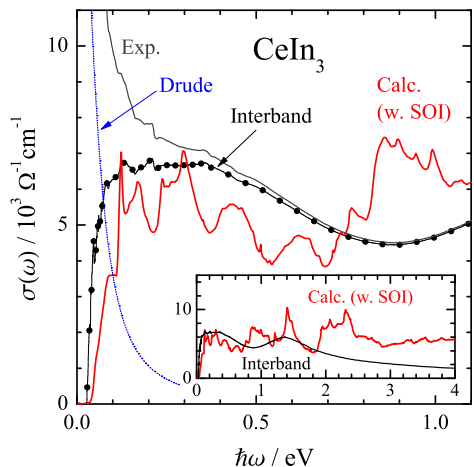}
\end{center}
\caption{
(Color online) Experimental and calculated optical conductivity [$\sigma(\omega)$] spectra of CeIn$_3$.
The calculation was performed with SOI.
The experimental spectrum was taken in the paramagnetic phase at $T$~=~60~K.
The interband spectrum was obtained by subtracting the Drude component due to the carriers from the experimental $\sigma(\omega)$ spectrum.
The Drude curve was evaluated at the slope at around 0.04~eV.
The inset shows the calculated $\sigma(\omega)$ spectrum compared with the experimental interband spectrum in the wide energy range below 4~eV.
}
\label{CeIn3OC}
\end{figure}
The calculated electronic structure shown in Fig.~\ref{CeIn3Band} is consistent with the previous work.~\cite{Lalic2001}
Figure~\ref{CeIn3OC} shows the calculated and experimental $\sigma(\omega)$ spectra of CeIn$_3$.
To make a comparison with the paramagnetic band calculation, the experimental spectrum was taken in the paramagnetic phase at $T$~=~60~K.
The interband part after subtracting the Drude curve, which was fitted at around 0.04~eV, from the experimental $\sigma(\omega)$ spectrum has a broad mid-IR peak at 0.2~eV.
The peak does not have the SOS structure, because the peak is broader than the splitting energy.
The calculated spectrum has some small peaks originating from the optical transition in some bands (bold lines in Fig.~\ref{CeIn3Band}) near $E_{\rm F}$.
Although the calculated spectrum has some small peaks, the overall spectral shape corresponds to the experimental spectrum below 0.8~eV, for instance the flat-top shape of the experimental spectrum from 0.1 to 0.4~eV.
Meanwhile, several peaks in the energy range of 0.9~--~1.9~eV in the calculation can be regarded to be equivalent to the broad peak at 1.4~eV in the experiment as shown in the inset of Fig.~\ref{CeIn3OC}, because the experimental life time broadening seems to be larger than that in the calculated spectrum in this energy region.
So the dip structure at around 0.6~eV in the calculated $\sigma(\omega)$ spectrum also corresponds to the minimum at 0.9~eV in the experimental spectrum.
The broad mid-IR peak of CeIn$_3$ can therefore also be concluded to originate from the band structure.
As mentioned earlier, the experimental spectrum in the antiferromagnetic phase at 8~K is identical to that in the paramagnetic phase at 60~K.
This result is also evidence that the mid-IR peak originates from the band structure.

%
So far, the experimental $\sigma(\omega)$ spectra of CePd$_3$, CeSn$_3$, and CeIn$_3$ have been demonstrated to be in good agreement with the spectra derived from the band structure calculation including SOI.
In addition, the triple-peak structure in CePd$_3$ can be explained with the SOS of the Ce~$4f$ state.
These results indicate that the mid-IR peaks of these materials fundamentally originate from the band structure and the SOS of the Ce~$4f$ state plays an important role in the shape of the mid-IR peaks.
This is consistent with the fact that the mid-IR peaks of many other Ce compounds have a double-peak structure with energy splitting of 0.25~eV.
The spectral shapes of CePd$_3$, CeSn$_3$, and CeIn$_3$ are significantly different from one another.
This indicates that the spectral shape originates from the particular electronic structure near $E_{\rm F}$ in each case.
Therefore, the mid-IR peaks can be explained by the band structure calculation with SOS and without reference to the $cf$ hybridization gap.
This result is reasonable in the mixed-valent Ce compounds, because the $4f$ states can be treated as itinerant bands, for instance, the band structure calculation of CeSn$_3$ can reproduce the angle-dependent frequency of the de Haas-van Alphen effect.~\cite{Hasegawa1990}

%
To summarize, we calculated the $\sigma(\omega)$ spectra of CePd$_3$, CeSn$_3$ and CeIn$_3$ in the infrared region from the first-principle band calculation with SOI and compared them with the experimental spectra.
The calculated $\sigma(\omega)$ spectra are consistent with the experimental spectra, even though the experimental $\sigma(\omega)$ spectra of these materials are significantly different from one another.
The different spectral shapes originate from the different electronic structures near $E_{\rm F}$.
These results indicate that the mid-IR peaks originate not from the conventional $cf$ hybridization gap but from the characteristic band structure.

\section*{Acknowledgments}
This work was a joint studies program of the Institute for Molecular Science (2007) and was partially supported by a Grant-in-Aid of Scientific Research (B) (No.~18340110) from MEXT of Japan.

%
\end{document}